\documentclass[12pt]{article}
\usepackage{axodraw}

\catcode`@=12
\begin{document}
\thispagestyle{empty}
\begin{flushright} 
UCRHEP-T388\\ 
May 2005\
\end{flushright}
\vspace{0.5in}
\begin{center}
{\LARGE	\bf Discrete Symmetry and $CP$ Phase\\ of the Quark Mixing Matrix\\}
\vspace{1.5in}
{\bf Shao-Long Chen and Ernest Ma\\}
\vspace{0.2in}
{\sl Physics Department, University of California, Riverside, 
California 92521\\}
\vspace{1.5in}
\end{center}
\begin{abstract}\
A simple specific pattern of the two $3 \times 3$ quark mass matrices is 
proposed, resulting in a prediction of the $CP$ phase of the charged-current 
mixing matrix $V_{CKM}$, i.e. $\sin 2 \phi_1 (\beta) = 0.733$, which is in 
remarkable agreement with data, i.e. $\sin 2 \phi_1 = 0.728 \pm 0.056 \pm 
0.023$ from Belle and $\sin 2 \beta = 0.722 \pm 0.040 \pm 0.023$ from Babar.  
This pattern can be maintained by a discrete family symmetry, an example 
of which is $D_7$, the symmetry group of the heptagon.
\end{abstract}

\newpage
\baselineskip 24pt

The three families of quarks have very different masses and mix with 
one another in the charged-current mixing matrix $V_{CKM}$ in a nontrivial 
manner.  This $3 \times 3$ matrix has three angles and one phase, the latter 
being the source of $CP$ nonconservation in the Standard Model (SM) of 
particle interactions.  In the context of the SM, this phase is now 
measured with some precision, i.e.
\begin{equation}
\sin 2 \phi_1 = 0.728 \pm 0.056 \pm 0.023
\end{equation}
from Belle \cite{belle}, and
\begin{equation}
\sin 2 \beta = 0.722 \pm 0.040 \pm 0.023
\end{equation}
from Babar \cite{babar}, where $\phi_1$ (also known as $\beta$) is defined 
as the phase of the element $V_{td}$, 
i.e.
\begin{equation}
V_{td} = |V_{td}| e^{-i \phi_1}.
\end{equation}
Together with $|V_{us}|$, $|V_{cb}|$, and $|V_{ub}|$, the entire $V_{CKM}$ 
matrix can now be fixed, up to sign and phase conventions.  Given the 
experimentally measured values of these parameters, is there a pattern 
to be recognized?  The answer is not obvious, because the relevant 
physics comes from the structure of the two $3 \times 3$ quark mass 
matrices
\begin{eqnarray}
{\cal M}_u &=& V^u_L \pmatrix{m_u & 0 & 0 \cr 0 & m_c & 0 \cr 0 & 0 & m_t} 
(V^u_R)^\dagger, \\ 
{\cal M}_d &=& V^d_L \pmatrix{m_d & 0 & 0 \cr 0 & m_s & 0 \cr 0 & 0 & m_b} 
(V^u_R)^\dagger,
\end{eqnarray}
from which the observed quark mixing matrix is obtained:
\begin{equation}
V_{CKM} = (V^u_L)^\dagger V_L^d.
\end{equation}

A theoretically consistent approach to understanding ${\cal M}_u$ and 
${\cal M}_d$ is to extend the Lagrangian of the SM to support a family 
symmetry in such a way that the forms of these mass matrices are 
restricted with fewer parameters than are observed, thus making 
one or more predictions.  Because of complex phases, this is often not 
a straightforward proposition.  In this paper, we advocate a simple specific 
pattern, i.e. ${\cal M}_u$ is diagonal, whereas ${\cal M}_d$ is of the 
form
\begin{equation}
{\cal M}_d = \pmatrix{0 & a & \xi b \cr a & 0 & b \cr \xi c & c & d},
\end{equation}
which was first proposed by one of us long ago \cite{m91}.  The difference 
here is that whereas $|\xi|$ was fixed at $m_u/m_c$ in that model, it is 
now a free parameter.  The family symmetry used previously was $S_3 \times 
Z_3$, which still works, but with different $Z_3$ assignments and a larger 
Higgs sector.  As a more elegant example for our discussion, we choose 
instead $D_7$, the symmetry group of the heptagon \cite{m04}.  A recent 
proposal \cite{bk05} based on $Q_6$ has both ${\cal M}_u$ and ${\cal M}_d$ 
of the form of Eq.~(7), but with $\xi=0$.  To maintain this latter condition 
consistently, an extra $Z_{12}$ symmetry has to be assumed.  Here $\xi$ is 
simply another parameter, equal to the ratio of two arbitrary vacuum 
expectation values.

The group $D_7$ has 14 elements, 5 equivalence classes, and 5 irreducible 
representations. Its character table is given by

\begin{table}[htb]
\caption{Character Table of $D_7$.}
\begin{center}
\begin{tabular}{|c|c|c|c|c|c|c|c|}
\hline
class & $n$ & $h$ & $\chi_1$ & $\chi_2$ & $\chi_3$ & $\chi_4$ & $\chi_5$ \\ 
\hline
$C_1$ & 1 & 1 & 1 & 1 & 2 & 2 & 2 \\
$C_2$ & 7 & 2 & 1 & $-1$ & 0 & 0 & 0 \\
$C_3$ & 2 & 7 & 1 & 1 & $a_1$ & $a_2$ & $a_3$ \\
$C_4$ & 2 & 7 & 1 & 1 & $a_2$ & $a_3$ & $a_1$ \\
$C_5$ & 2 & 7 & 1 & 1 & $a_3$ & $a_1$ & $a_2$ \\
\hline
\end{tabular}
\end{center}
\end{table}

Here $n$ is the number of elements and $h$ is the order of each element. 
The numbers $a_k$ are given by $a_k = 2 \cos (2 k \pi/7)$.  The character 
of each representation is its trace and must satisfy the 
following two orthogonality conditions:
\begin{eqnarray}
\sum_{C_i} n_i \chi_{ai} \chi^*_{bi} = n \delta_{ab}, ~~~~~ 
\sum_{\chi_a} n_i \chi_{ai} \chi^*_{aj} = n \delta_{ij},
\end{eqnarray}
where $n = \sum_i n_i$ is the total number of elements.  The number of 
irreducible representations must be equal to the number of equivalence classes.

The three irreducible two-dimensional representations of $D_7$ may be chosen 
as follows.  For {\bf 2}$_1$, let
\begin{eqnarray}
&& C_1: \pmatrix{1 & 0 \cr 0 & 1}, ~~~ C_2: \pmatrix{0 & \omega^k \cr 
\omega^{7-k} & 0}, ~(k=0,1,2,3,4,5,6), \nonumber \\ 
&& C_3: \pmatrix{\omega & 0 \cr 0 & \omega^6}, ~\pmatrix{\omega^6 & 0 \cr 0 
& \omega}, ~~~ C_4: \pmatrix{\omega^2 & 0 \cr 0 & \omega^5}, ~\pmatrix{
\omega^5 & 0 \cr 0 & \omega^2}, \nonumber \\  && C_5: \pmatrix{\omega^4 & 0 
\cr 0 & \omega^3}, ~\pmatrix{\omega^3 & 0 \cr 0 & \omega^4},
\end{eqnarray}
where $\omega = \exp(2 \pi i/7)$, then {\bf 2}$_{2,3}$ are simply obtained by 
the cyclic permutation of $C_{3,4,5}$.

For $D_n$ with $n$ prime, there are $2n$ elements divided into $(n+3)/2$ 
equivalence classes: $C_1$ contains just the identity, $C_2$ has the $n$ 
reflections, $C_k$ from $k = 3$ to $(n+3)/2$ has 2 elements each of order 
$n$.  There are 2 one-dimensional representations and $(n-1)/2$ 
two-dimensional ones. For $D_3=S_3$, the above reduces to the ``complex'' 
representation with $\omega=\exp(2 \pi i/3)$ discussed in a recent review 
\cite{fuji}.

The group multiplication rules of $D_7$ are:
\begin{eqnarray}
&& {\bf 1'} \times {\bf 1'} = {\bf 1}, ~~~ {\bf 1'} \times {\bf 2}_i = 
{\bf 2}_i, \\
&& {\bf 2}_i \times {\bf 2}_i = {\bf 1} + {\bf 1'} + {\bf 2}_{i+1}, ~~~ 
{\bf 2}_i \times {\bf 2}_{i+1} = {\bf 2}_i + {\bf 2}_{i+2},
\end{eqnarray}
where {\bf 2}$_{4,5}$ means {\bf 2}$_{1,2}$.  In particular, let $(a_1,a_2), 
(b_1,b_2) \sim {\bf 2}_1$, then
\begin{equation}
a_1 b_2 + a_2 b_1 \sim {\bf 1}, ~~~  a_1 b_2 - a_2 b_1 \sim {\bf 1'}, ~~~  
(a_1 b_1, a_2 b_2) \sim {\bf 2}_2.
\end{equation}
In the decomposition of ${\bf 2}_1 \times {\bf 2}_2$, we have instead
\begin{equation}
(a_2 b_1, a_1 b_2) \sim {\bf 2}_1, ~~~(a_2 b_2, a_1 b_1) \sim {\bf 2}_3.
\end{equation}

To arrive at our proposed pattern, let
\begin{eqnarray}
&& (u,d)_i \sim {\bf 2}_1 + {\bf 1}, ~~~d^c_i \sim {\bf 2}_1 + {\bf 1}, 
~~~u^c_i \sim {\bf 2}_2 + {\bf 1}, \\
&& \phi^d_i \sim {\bf 2}_1 + {\bf 1}, ~~~\phi^u_i \sim {\bf 2}_3 + {\bf 1},
\end{eqnarray}
where the scalar fields $\phi^{d,u}_i$ are distinguished by an extra 
symmetry such as supersymmetry so that they couple only to $d^c,u^c$ 
respectively.  Using the multiplication rules listed above, we see 
that ${\cal M}_u$ is indeed diagonal, and ${\cal M}_d$ is of the form 
of Eq.~(7), with $a,d$ coming from $\langle \phi^d_3 \rangle$ and 
$(b,\xi b)$, $(c,\xi c)$ from $\langle \phi^d_{1,2} \rangle$ respectively.  
These latter are distinct from $\langle \phi^u_{1,2} \rangle$, so that the 
constraint $|\xi|=m_u/m_c$ in Ref.~\cite{m91}  no longer applies.

As in Ref.~\cite{m91}, we can redefine the phases of ${\cal M}_d$ so that 
$a,b,c,d$ are real, but $\xi$ is complex.  Assuming that $a^2 << b^2$ and 
$|\xi|^2 << 1$, then to a very good approximation,
\begin{eqnarray}
&& m_b \simeq \sqrt{c^2+d^2}, ~~~ m_s \simeq {bc \over \sqrt{c^2+d^2}}, 
~~~ m_d \simeq \left| {a^2 d \over bc} - 2 \xi a \right|, \\ 
&& V_{cb} \simeq {bd \over c^2+d^2}, ~~~ V_{us} \simeq -{ad \over bc} + \xi, 
~~~ V_{ub} \simeq {ac+\xi bd \over c^2+d^2}.
\end{eqnarray}
Using the 6 experimental inputs on $m_b$, $m_s$, $m_d$, $|V_{cb}|$, 
$|V_{us}|$, and $|V_{ub}|$, the 6 parameters $a$, $b$, $c$, $d$, $Re\xi$, 
and $Im\xi$ are fixed, thereby predicting the $CP$ phase of $V_{CKM}$.
Numerical inputs of quark masses (in GeV) are taken from Ref.~\cite{qm} 
evaluated at the scale $M_W$, i.e.
\begin{equation}
m_d = 0.00473 \pmatrix{+0.00061 \cr -0.00067}, ~~~ m_s = 0.0942 
\pmatrix{+0.0119 \cr -0.0131}, ~~~ m_b = 3.03 \pm 0.11.
\end{equation}
Numerical inputs of mixing angles are taken from the 2004 Particle Data 
Group compilation \cite{pdg}, i.e.
\begin{equation}
|V_{us}| = 0.2200 \pm 0.0026, ~~~ |V_{cb}| = (41.3 \pm 1.5) \times 10^{-3}, 
~~~ |V_{ub}| = (3.67 \pm 0.47) \times 10^{-3}.
\end{equation}
Taking the central values of the above 6 quantities, we find
\begin{eqnarray}
&& a=0.0142~{\rm GeV}, ~~b=0.1566~{\rm GeV}, ~~c=1.8223~{\rm GeV}, 
~~d=-2.4208~{\rm GeV}, \\
&& Re \xi = 0.08124, ~~~ Im \xi = 0.08791.
\end{eqnarray}
After rotating the phase of $V_{us}$ to make it real to conform to the 
standard convention, we then predict
\begin{equation}
\sin 2 \phi_1 = 0.733,
\end{equation}
in remarkable agreement with experiment, i.e. Eqs.~(1) and (2).  The three 
angles $\phi_1 (\beta)$, $\phi_2 (\alpha)$, $\phi_3 (\gamma)$ of the unitarity 
triangle are then $23.6^\circ, 98.4^\circ, 58.0^\circ$ respectively.

We may also vary the 6 numerical inputs within their allowed ranges, taking 
into account the correlation between $m_d$ and $m_s$ (because $m_d/m_s$ 
is tightly constrained.) In that case,
\begin{equation}
\sin 2 \phi_1 = 0.733 \pmatrix{+0.107 \cr -0.152}.
\end{equation}
In the future, these input parameters will be determined with more 
precision and our model will be more severely tested.

Flavor-changing neutral-current interactions are mediated by the three 
neutral Higgs bosons in the $d$ sector with Yuakawa couplings given by
\begin{eqnarray}
{\cal L}_Y &=& {a \over v_3} \phi^0_3 (q_1 q^c_2 + q_2 q^c_1) + {b \over v_1} 
(\phi^0_1 q_2 + \phi^0_2 q_1) q^c_3 \nonumber \\ 
&+& {c \over v_1} q_3 (\phi^0_1 q^c_2 + \phi^0_2 q^c_1) + {d \over v_3} 
\phi^0_3 q_3 q^c_3 + h.c.,
\end{eqnarray}
where $q_i, q^c_j$ are the basis states of the mass 
matrix ${\cal M}_d$ of Eq.~(7).  Let
\begin{equation}
{\cal M}_d = V \pmatrix{m_d & 0 & 0 \cr 0 & m_s & 0 \cr 0 & 0 & m_b} 
(V^c)^\dagger,
\end{equation}
then $V=V_{CKM}$ and $V^c$ is its analog for the charge-conjugate states. 
In this model, they are approximately given by
\begin{equation}
V \simeq \pmatrix{1 & -(ad/bc)+\xi & (ac+\xi bd)/(c^2+d^2) \cr (ad/bc)-\xi^* 
& 1 & bd/(c^2+d^2) \cr -a/c & -bd/(c^2+d^2) & 1},
\end{equation}
where $\xi = v_2/v_1$, and
\begin{equation}
V^c \simeq \pmatrix{1 & -(ad/bc)+\xi^* c^2/(c^2+d^2) & a/b + \xi^* cd/
(c^2+d^2) \cr a \sqrt{c^2+d^2}/bc & d/\sqrt{c^2+d^2} & -c/\sqrt{c^2+d^2} 
\cr -\xi c/\sqrt{c^2+d^2} & c/\sqrt{c^2+d^2} & d/\sqrt{c^2+d^2}}.
\end{equation}
Using $q_i = V_{i\alpha} d_\alpha$ and $q^c_j = V^c_{j\beta} d^c_\beta$, 
we can rewrite the couplings of $\phi^0_{1,2,3}$ in terms of the quark mass 
eigenstates and evaluate their contributions to flavor-changing processes 
such as $K-\overline{K}$ and $B-\overline{B}$ mixings, etc.

An important point to notice \cite{m01} is that if $\phi_{1,2}$ are 
replaced by $\phi_3$ in the Yukawa sector, then there would be no 
flavor-changing interactions at all.  Hence all such effects are contained 
in the terms
\begin{equation}
\left( {\phi^0_1 \over v_1} - {\phi^0_3 \over v_3} \right) (b q_2 q^c_3 
+ c q_3 q^c_2) + \xi \left( {\phi^0_2 \over v_2} - {\phi^0_3 \over v_3} 
\right) (b q_1 q^c_3 + c q_3 q^c_1).
\end{equation}
Whereas the mass of the SM combination $(v_1 \phi^0_1 + v_2 
\phi^0_2 + v_3 \phi^0_3)/\sqrt{|v_1|^2+|v_2|^2+|v_3|^2}$ should be of order 
the electroweak breaking scale, the two orthogonal combinations contained 
in the above are allowed to be much heavier, say a few TeV.

The $K_L - K_S$ mass difference gets its main contribution from the 
$(q_1 q^c_3)(q_3 q^c_1)^\dagger$ term through $\phi^0_2$ exchange.  Thus
\begin{equation}
{\Delta m_K \over m_K} \simeq {B_K f_K^2 b^2 c^2 d \over 3  
(c^2 + d^2)^{3/2} m_2^2 v_1^2}.
\end{equation}
Using $f_K = 114$ MeV, $B_K = 0.4$, $v_1 = 100$ GeV, and $m_2 = 7$ TeV, 
we find this contribution to be $2.5 \times 10^{-17}$, well below the 
experimental value of $7.0 \times 10^{-15}$.  Similarly,
\begin{equation}
{\Delta m_B \over m_B} \simeq {B_B f_B^2 b c d \over 3 
(c^2 + d^2)^{1/2} m_2^2 v_1^2},
\end{equation}
and
\begin{equation}
{\Delta m_{B_s} \over m_{B_s}} \simeq {B_B f_B^2 b c d^2 \over 3  
(c^2 + d^2) m_1^2} \left( {1 \over v_1^2} + {1 \over v_3^2} \right).
\end{equation}
Using $f_B = 170$ MeV, $B_B = 1.0$, $v_{1,3} = 100$ GeV, and $m_{1,2} = 7$ 
TeV, we find these contributions to be $4.5 \times 10^{-15}$ and $7.2 \times 
10^{-15}$ respectively, to be compared against the experimental value of 
$6.3 \times 10^{-14}$ for the former and the experimental lower bound of 
$1.8 \times 10^{-12}$ for the latter.

It is interesting to note that the form of Eq.~(7) is easily adaptable to 
the Majorana neutrino mass matrix.  By rearranging the two zeros, we 
can have
\begin{equation}
{\cal M}_\nu^{(e, \mu, \tau)} = \pmatrix{a & c & d \cr c & 0 & b \cr d & b & 0}
\end{equation}
as advocated in Ref.~\cite{m04}, which is a successful description of 
neutrino oscillation phenomena.  This hints at the intriguing possibility 
that despite their outward dissimilarities, both quark and lepton family 
structures may actually come from the same mold.

In conclusion, we have pointed out that the ${\cal M}_d$ of Eq.~(7) predicts 
the correct value of the $CP$ phase of the quark mixing matrix.  Its form 
is derivable from a discrete family symmetry such as $D_7$, which also 
works for leptons as previously shown.  Extra Higgs doublets are predicted, 
but their contributions to flavor-changing interactions are suitably 
suppressed if their masses are of order a few TeV.

This work was supported in part by the U.~S.~Department of Energy
under Grant No.~DE-FG03-94ER40837.

\newpage
\bibliographystyle{unsrt}

\end{document}